\documentclass[
amsmath, %
floatfix, %
twocolumn, %
showpacs, %
reprint, %
prb, %
aps, %
nofootinbib,%
nobibnotes,%
]{revtex4-1}
\usepackage[T1]{fontenc}
\usepackage{graphicx}
\usepackage{latexsym,amsthm,dsfont,xcolor}

\usepackage[looser]{newtxtext}
\usepackage[subscriptcorrection,nosymbolsc,smallerops,bigdelims]{newtxmath}
\DeclareMathAlphabet{\mathcal}{OMS}{cmsy}{m}{n}
\usepackage{bm}

\usepackage{floatrow}
\usepackage[hyperindex,breaklinks]{hyperref}
\hypersetup{
	colorlinks,
	linkcolor={blue!100!black},
	citecolor={blue!100!black},
	urlcolor={blue!100!black}
}
\makeatletter
\renewcommand*{\eqref}[1]{%
	\hyperref[{#1}]{\textup{\tagform@{\ref*{#1}}}}%
}
\makeatother

\definecolor{light-gray}{gray}{0.95}

\newcommand{\bk}{b_{\bm{k},\lambda}}
\newcommand{\bkd}{b_{\bm{k},\lambda}^{\dag}}
\newcommand{\bmkd}{b_{-\bm{k},\lambda}^{\dag}}
\newcommand{\sumk}{\sum_{\bm{k},\lambda}}

\newcommand{\wk}{\omega_{\bm{k},\lambda}}

\newcommand{\fk}{f_{\bm{k},\lambda}}

\newcommand{\ket}[1]{\left| #1 \right>} 
\newcommand{\bra}[1]{\left< #1 \right|}

\newcommand{\be}{\begin{equation}}
\newcommand{\ee}{\end{equation}}
\newcommand{\bea}{\begin{eqnarray}}
\newcommand{\eea}{\end{eqnarray}}
\newcommand{\bdm}{\begin{displaymath}}
\newcommand{\edm}{\end{displaymath}}
\newcommand{\hf}{\frac12}

\newcommand{\mev}{\mathrm{meV}}

\renewcommand\Re{\operatorname{Re}}

\newcommand{\upemhf}{\vspace{-0.7em}}

\begin{document}
\author{Micha{\l} Gawe{\l}czyk}\email{michal.gawelczyk@pwr.edu.pl}\affiliation{Institute of Physics, Wroc{\l}aw University of
Technology, 50-370 Wroc{\l}aw, Poland}
\author{Pawe{\l} Machnikowski}\affiliation{Institute of Physics, Wroc{\l}aw University of
Technology, 50-370 Wroc{\l}aw, Poland}
\title{Optical initialization of hole spins in p-doped quantum dots: orientation efficiency and loss of coherence}

\begin{abstract}
We study theoretically the recently proposed hole spin initialization scheme for p-doped quantum well or dot systems via coupling to trion states with sub-picosecond circularly polarized laser pulses. We analyze the efficiency of spin initialization an predict the intrinsic spin coherence loss due to the pulse excitation itself as well as the phonon-induced spin dephasing, both taking place on the timescale of the driving laser pulse. We show that the ratio of the degree of dephasing to the achieved orientation effect does not depend on the pulse area but is sensitive to the temperature and detuning. The optimal excitation parameters are identified.
\end{abstract}

\maketitle

%#############################################################################
\section{Introduction}\label{intro}
Exploration of spin dynamics and spin decoherence in semiconductors is of a great
interest in view of possible applications in the rapidly expanding field of spintronics and can contribute to the development of new devices, like logic circuits\cite{Datta, Chiolerio}, magnetic random access memories\cite{Gallagher} or spin-transfer nano-oscillators\cite{Kiselev}. Despite the fact that hole spin states received less attention in the past due to their sub-picosecond dephasing time in a bulk GaAs\cite{Hilton, Shen} they turn out to be far more promising candidates for the spin control in nanostructure-based devices. Firstly, the reduction of the system dimensionality yields the coherence time extension up to a few picoseconds in p-doped quantum well (QW) systems\cite{Damen, Schneider, Lu} which can still be enhanced by a further carrier localization at low temperatures\cite{Saper, Korn1, Korn2, Kugler, Studer}. Moreover, in cotradiction to electron states, holes do not suffer from the spin dephasing induced by the contact hyperfine interaction\cite{Fischer}. Furthermore, due to the high anisotropy of the hole g-factor in low-dimensional GaAs, new possibilities of spin manipulation arise\cite{Machnikowski, Andlauer, Kugler2}.

Recent investigations of spin dynamics of resident holes in p-modulation-doped QWs came with a proposal for an optical spin arrangement scheme\cite{Saper,Korn1,Korn2,Kugler,Studer}. The system is driven by a circularly polarized, sub-picosecond laser pulse which excites part of resident holes confined in QW fluctuations to the trion state. As the light is circularly polarized it couples only electrons and holes with spins oriented in one direction. The system is placed in a homogenous in-plane magnetic field and therefore the excited trion spin undergoes precession and, as a consequence, recombines with a hole in a random spin state. Thus, on the average, half of the created trions leave an inverted hole spin upon recombination. In this manner, one is able to polarize hole spins in a desired direction.

Our aim is to investigate this spin initialization scheme in terms of its efficiency as well as of the intrinsic (caused by the excitation process itself) and phonon-induced hole spin dephasing. Such decay of coherence may be important in future applications but it affects also the results of current experiments\cite{VarvigArXiv}, in particular those based on the resonant spin amplification effect\cite{Kikkawa, Kugler, Varwig}, where the spin coherence is crucial for the formation of the observed signal.

It is known\cite{GrodeckaPRA} that under optical excitation spin states can experience pure dephasing due to a dynamical phonon response to the transient charge evolution. The spin state, even if not directly coupled to the reservoir, is affected by an indirect dephasing through the entangling charge evolution caused by the optical pulse. In this paper, we show that, indeed, such a hole spin dephasing process, resulting from the carrier-phonon interaction, is present in the discussed system. Moreover, we find that the laser pulse itself causes a considerable amount of decoherence. Thus, we show that some degree of spin dephasing is unavoidably built into the initialization scheme.

Both these dephasing processes are of a dynamical nature and as such cannot be described as a simple exponential temporal decay of coherence characterized by a dephasing time. To compare it to other processes occurring in systems of this type one can use the only characteristics available, namely the total decrease of coherence and the duration of dephasing process which is equal to the laser pulse duration (here, sub-picosecond). Comparing the latter to timescales of other relevant processes occurring in discussed system, like the phonon-assisted\cite{Trif, Fras} or hyperfine hole-nuclei coupling driven hole spin relaxation\cite{Fras2}, electron spin dephasing\cite{Masumoto, Saper2, Hernandez} or relaxation of the positive trion\cite{Braun} (ranging from hundreds of picoseconds to a few microseconds), we find it to be at least three orders of magnitude shorter. This fact allows us to consider it separately and treat the dynamical spin dephasing as instantaneous on the background of the whole optical spin initialization process. One can therefore think of the dynamically dephased hole spin state as of the initial state for all other processes occurring in the discussed spin orientation scheme\cite{Kugler}.

In this paper, we discuss the dependence of the efficiency of spin polarization and the degree of spin dephasing on both carrier system and laser pulse features. We show that the degree of dynamical dephasing can be on the order of a few to even tens of percent, depending on the system features, which makes it highly relevant in comparison with other unfavourable processes affecting the quality of spin initialization scheme. We show that the relative magnitude of the phonon-induced effect, as compared to the intrinsic dephasing and spin orientation efficiency, changes qualitatively between low and moderate temperatures and indicate the optimal control conditions for minimizing the loss of coherence.

The paper is organized as follows. In Sec.~\ref{model}, we provide the theoretical framework of the investigated system. Then, in Sec~\ref{spin} we analyze the spin initialization process and discuss its efficiency. Next, in Sec.~\ref{Intrinsic}, we investigate the intrinsic loss of hole spin coherence due to the optical excitation and recombination process. Then, in Sec.~\ref{Phonon} we study the phonon-induced hole spin dephasing. The results of the latter are given in Sec.~\ref{results}. Finally, we conclude the paper in Sec.~\ref{conclusions}.

%#####################################################################
\section{Model}\label{model}
We investigate a single p-doped quantum dot (QD), which can be both a self-assembled QD or a width fluctuation in a single p-doped QW, forming a so called natural or monolayer fluctuation QD\cite{Gammon}, with a resident hole present in it. This system is optically excited by a short (sub-picosecond) pulse of circularly polarized laser light. The carrier system is coupled to a bulk acoustic phonon bath by means of deformation potential and piezoelectric couplings (the Fr{\"{o}}hling coupling is not considered as only low frequencies are relevant under the excitation conditions considered here). The Hamiltonian of the system thus has the form
\be\label{hamiltonian}
H=H_\mathrm{c}+H_{\mathrm{ph}}+H_{\mathrm{c-ph}}+H_{\mathrm{las}}\mathrm{.}\nonumber
\ee
The first term, the Hamiltonian of the confined carrier subsystem, is
\be\label{h_c}
H_\mathrm{c}=E_\mathrm{t}\ket{T\!\uparrow}\!\!\bra{T\!\uparrow}\mathrm{,}\nonumber
\ee
with only one relevant trion state taken into account (according to the selection rules) and the energies of the hole states $\ket{\uparrow}$, $\ket{\downarrow}$ set to zero. Here, $\ket{\uparrow}=h_{\uparrow}^\dagger \ket{0}$, $\ket{\downarrow}=h_{\downarrow}^\dagger \ket{0}$ and $\ket{T\!\uparrow}=h_{\uparrow}^\dagger h_{\downarrow}^\dagger a_{\uparrow}^\dagger \ket{0}$, where $h_{\downarrow\slash\uparrow}^\dagger$ and $a_{\downarrow\slash\uparrow}^\dagger$ are the hole and electron creation operators with the spin orientation denoted in the subscript, and $E_\mathrm{t}$ is the trion energy.

The circularly polarized pumping laser couples only the two optically active states, which is reflected in the laser Hamiltonian part
\be\label{h_las}
H_{\mathrm{las}}=\hf f\left( t \right) \mathrm{e}^{i\Delta t} \ket{\uparrow}\!\!\bra{T\!\uparrow} + \mathrm{h.c.} \mathrm{,}\nonumber
\ee 
where $f\left({t}\right)$ is the laser pulse envelope, taken to be Gaussian,
\be\label{envelope}
f\left( t\right)=\frac{\theta}{\sqrt{2\pi}\tau}\exp{\left(- \frac{t^2}{2\tau^2} \right)},\nonumber
\ee
with the pulse duration time $\tau=600~\mathrm{fs}$ and the pulse area $\theta=\int_{-\infty}^{\infty}{f\left(t\right)\mathrm{d}t}$. $\Delta$ is the frequency detuning from the fundamental transition in the system and we use the rotating wave approximation and the rotating frame picture\cite{Param}.

\begin{table}
	\begin{tabular*}{\columnwidth}{@{\extracolsep{\fill}}lll}
		\toprule
		Static dielectric constant & $\varepsilon_{\mathrm{s}}$ & 12.9 \\
		Piezoelectric constant & $d$ & -0.16 C/m$^{2}$ \\
		Longitudinal sound speed & $c_{\mathrm{l}}$ & 5150 m/s \\
		Transverse sound speed & $c_{\mathrm{t1, t2}}$ & 2800 m/s \\
		Deformation potential: & & \\ 
		\hspace{0.7em} for electrons & $\sigma_\mathrm{e}$ & $-7.0$ eV \\
		\hspace{0.7em} for holes & $\sigma_\mathrm{h}$ & $3.5$ eV \\
		Crystal density & $\rho_\mathrm{c}$ & 5350 kg/m$^{3}$   \\
		\toprule
	\end{tabular*}
	\caption{\label{tab:param}The GaAs material parameters used in the calculations.\upemhf}
\end{table}
The system is then coupled to the bulk acoustic phonon reservoir (which is justified by the similarity of the elastic properties of the surrounding and the QW material). The phonons are described by the free phonon Hamiltonian,
\be 
H_{\mathrm{ph}}=\sumk{\hbar\wk\bkd\bk},\nonumber 
\ee
and coupled to the carrier subsystem by the interaction Hamiltonian $H_{\mathrm{c-ph}}$
\begin{align}
H_{\mathrm{c-ph}}=&\sumk \left[ F_\mathrm{h}\left( \bm{k},\lambda \right)\left(\ket{\downarrow}\!\!\bra{\downarrow}+\ket{\uparrow}\!\!\bra{\uparrow}\right)\right. \nonumber
\\ &\left.+F_\mathrm{t}\left( \bm{k},\lambda \right)\ket{T\!\uparrow}\!\!\bra{T\!\uparrow} \right]\left( b_{\bm{k}} + b_{- {\bm{k}}}^\dagger \right)\mathrm{,}\nonumber
\end{align}
where $\bkd$ and $\bk$ are phonon creation and annihilation operators, $\wk$ is the phonon frequency and $\lambda$ denotes the phonon branch (denoted $\mathrm{l}$ and $\mathrm{t1}$, $\mathrm{t2}$ for the longitudinal and the two transverse branches, respectively). The coupling constants $F_\mathrm{h}\left( \bm{k},\lambda \right)$ and $F_\mathrm{t}\left( \bm{k},\lambda \right)=2F_\mathrm{h}\left( \bm{k},\lambda \right)-F_\mathrm{e}\left( \bm{k},\lambda \right)$ include the deformation potential and piezoelectric couplings,
\begin{multline*}
F_\mathrm{e/h}{}\left( \bm{k},\lambda \right){}= \\ 
\sqrt{{}\frac{\hbar}{2NV{}\rho_\mathrm{c}}} {}\left({}\frac{id{e}M_\lambda{}\left(\bm{k}\right)}{\varepsilon_0\varepsilon_\mathrm{s}\sqrt{\wk}} {}+{}\delta_{\lambda\mathrm{l}}\sigma_\mathrm{e/h} \sqrt{{}\frac{k}{c_\mathrm{l}}}\right) {}\mathcal{F}_\mathrm{e/h}{}\left(\bm{k}\right),
\end{multline*}
where $d$ is the piezoelectric constant, $\varepsilon_\mathrm{s}$ is the static dielectric constant, $\sigma$ is the deformation potential, $\rho_\mathrm{c}$ is the material density, $c_\lambda$ is the speed of sound, $V$ is the unit cell volume, $N$ is the normalization constant, $\delta_{ij}$ is the Kronecker delta,
\be
\mathcal{F}_\mathrm{e/h}= \int_{-\infty}^{\infty}\mathrm{d}^3\bm{r}\psi^{*}_\mathrm{e/h}\left(\vec{r}\right)e^{i\bm{k}\bm{r}} \psi_\mathrm{e/h}\left(\vec{r}\right)\nonumber
\ee
is the formfactor\cite{Grodecka} and $M_\lambda\left(\bm{k}\right)$ are 
\begin{align*}
M_{\mathrm{l}}(\theta,\phi)  &=\frac{3}{2}\sin2\theta\cos\theta\sin2\phi, \\
M_{\mathrm{t1}}(\theta,\phi) &=\sin2\theta\cos2\phi, \\
M_{\mathrm{t2}}(\theta,\phi) &=(3\sin^{2}\theta-1)\cos\theta\sin 2\phi
\end{align*}
in the phonon polarization basis
\begin{align*}
\hat{e}_{\mathrm{l},\bm{k}} & = (\cos\theta\cos\phi,\cos\theta\sin\phi,\sin\theta),\\
    \hat{e}_{\mathrm{t1},\bm{k}} & = (-\sin\phi,\cos\phi,0),\\
    \hat{e}_{\mathrm{t2},\bm{k}} & = (\sin\theta\cos\phi,\sin\theta\sin\phi,-\cos\theta).
\end{align*}
The parameter values are presented in Tab.~\ref{tab:param}.

The Hamiltonian of the system under consideration in the absence of the laser field can be exactly diagonalized by application of unitary Weyl operators performing the transition into the polaron picture\cite{Haken, Roszak}
\be\label{weyl}
W_{\mathrm{h/t}}=\exp{\left[\sumk{\left(\frac{ F_{{\mathrm{h/t}}}\left( \bm{k},\lambda \right)  }{\hbar\wk} \bkd - \frac{ F_{{\mathrm{h/t}}}^{*}\left( \bm{k},\lambda \right)}{\hbar\wk} \bk \right)}\right]}.\nonumber
\ee 
The transformation operator $\mathds{W}$ for the two level system $\ket{\uparrow}$, $\ket{T\!\uparrow}$ is constructed as
\be\label{weylbig}
\mathds{W}=\left(\ket{\uparrow}\!\!\bra{\uparrow}+\ket{\downarrow}\!\!\bra{\downarrow}\right)W_\mathrm{h}+ \ket{T\!\uparrow}\!\!\bra{T\!\uparrow}W_\mathrm{t}\mathrm{.}\nonumber
\ee
Application of the $\mathds{W}$ operator on the full optically driven system cancels the original carrier-phonon interaction, $H_\mathrm{c-ph}$, and yields the transformed laser Hamiltonian part
\be\label{h_las_transform}
\tilde{H}_{\mathrm{las}}=\mathds{W}H_{\mathrm{las}}\mathds{W}^\dagger \simeq H_{\mathrm{las}}+V\mathrm{,}~\quad V=S\otimes R\mathrm{,}
\ee
where
\be\label{s}
S=-{\frac{i}{2}} f \left( t \right) \ket{\uparrow}\!\!\bra{T\!\uparrow} \mathrm{e}^{i \Delta t}+\mathrm{h.c.}\mathrm{,}
\ee\vspace{-15pt}
\be\label{r}
R=i \sumk{{\frac{\fk^{*}} {\hbar \wk}} \left( \bk - \bmkd \right) }\mathrm{.}
\ee
Here $\fk=\left[ {F_h\left(\bm{k},\lambda\right)-F_t\left(\bm{k},\lambda\right)} \right] /{\hbar\wk}$ and only terms up to the linear order in $\fk$ are kept in Eq.~\eqref{h_las_transform}. In this representation, there is no direct carrier-phonon interaction in place of which the phonon-assisted hole-trion optical transitions appear explicitly. These terms induce phonon response to spin-dependent charge evolution and lead to dephasing of the spin superpositions, as will be described in detail in Sec.~\ref{Phonon}.

%#####################################################
\section{Spin orientation}\label{spin}
Before we discuss the magnitude of the dephasing effects, we need a quantitative characteristics of the spin orientation itself. This is derived in this section.

The evolution of the density matrix of the carrier system under the optical excitation in the absence of phonons can be described perturbatively in the second order Born approximation in the interaction picture. The final state is
\begin{align}\label{ewolucjaimpuls}
\rho\left(\infty\right)\simeq&{\rho}_\mathrm{0}-\frac{i}{\hbar}\int_{-\infty}^{\infty}\mathrm{d}\tau \left[H_\mathrm{las}\left(\tau\right),\rho_0\right] \\
&-\frac{1}{\hbar^2}\int_{-\infty}^{\infty}\mathrm{d}\tau \int_{-\infty}^{\tau}\mathrm{d}\tau{'} \left[H_\mathrm{las}\left(\tau{}\right), \left[H_\mathrm{las}\left(\tau{'}\right),\rho_0\right]\right],\nonumber
\end{align}
where ${\rho}_\mathrm{0}$ is the initial density matrix with arbitrary occupations $p_{\uparrow}=p_{\uparrow}^{(0)}$, $p_{\downarrow}=p_{\downarrow}^{(0)}$ for spin-up and spin-down holes, respectively, and $p_{\mathrm{T}}=0$ for spin-up trions. Immediately after the excitation the occupations are $p_{\uparrow}=p_{\uparrow}^{(0)}\left(1-2q\right)$, $p_{\downarrow}=p_{\downarrow}^{(0)}$, $p_{\mathrm{T}}=2qp_{\uparrow}^{(0)}$, where
\be\label{q}
q=\frac{1}{8\hbar^2}{\left|\hat{f}\left(\Delta\right) \right|}^2 
\ee 
and
\be 
\hat{f}_t\left(\omega\right)=\int_{-\infty}^{t}\mathrm{d}\tau f\left(\tau\right)e^{i\omega{\tau}}.\nonumber
\ee
Thus, $q$ is proportional to the pulse power at the frequency $\omega=\Delta$.
Due to the in-plane magnetic field, the trion spin undergoes precession and recombines with a random hole spin. Therefore, on the average half of the trions leave an inverted hole spin after recombination. This yields the final relaxed state with no trions present and hole occupations equal to $p_{\uparrow}=p_{\uparrow}^{(0)}\left(1-q\right)$ and $p_{\downarrow}=p_{\downarrow}^{(0)}\left(1-q\right)+q$.

We introduce the spin polarization $P$ for the final state
\be
P=\frac{p_{\downarrow}-p_{\uparrow}}{p_{\uparrow}+p_{\downarrow}}=P_0+q\left(1-P_0\right),\nonumber
\ee 
where $P_0=p_{\downarrow}^{(0)}-p_{\uparrow}^{(0)}$ is the initial spin polarization. It is clear that $q=1$ corresponds to the full spin polarization, hence $q$ can be interpreted as the optical spin orientation efficiency.
Through its dependence on the pulse power, $q$ is proportional to $\theta^2$.

There are two contributions to the loss of spin coherence during the optical orientation process. One, which we refer to as intrinsic, results from the optical excitation and recombination, that is, it is related to the orientation mechanism itself. The other contribution is an effect of the coupling to phonons. These two contributions are discussed in Secs.~\ref{Intrinsic} and \ref{Phonon}, respectively. 
 
%#####################################################################
\section{Intrinsic decoherence}\label{Intrinsic}
Here, we calculate the degree of the intrinsic loss of coherence for the discussed spin initialization scheme, namely, the change in coherence due to the laser pulse itself without presence of phonons. Again, we use the second order Born approximation for the density matrix evolution. The object of our interest is the absolute value of the final hole spin coherence. From Eq.~\eqref{ewolucjaimpuls} one finds
\begin{align}
\left|\bra{\downarrow}\rho\left(\infty\right)\ket{\uparrow}\right|=& \left|\left(1-\xi\right)\bra{\downarrow}\rho_0\ket{\uparrow}\right|\nonumber\\
\simeq&\left(1-\Re\xi\right)\left|\bra{\downarrow}\rho_0\ket{\uparrow}\right|\nonumber,
\end{align}
where 
\be
\xi=\frac{1}{4\hbar^2}\int_{-\infty}^\infty\mathrm{d}\tau \int_{-\infty}^\tau\mathrm{d}\tau{'} {f\left(\tau\right)f\left(\tau{'}\right) e^{i\Delta\left(\tau-\tau{'}\right)}}.\nonumber
\ee
The real part of $\xi$ is
\begin{align}
\Re\xi&=\frac{1}{8\hbar^2}\int_{-\infty}^\infty\mathrm{d}\tau \int_{-\infty}^\tau\mathrm{d}\tau{'} {f\left(\tau\right)f\left(\tau{'}\right) \cos{\left(i\Delta\left(\tau-\tau{'}\right)\right)}}\nonumber\\
&=\frac{1}{8\hbar^2}\left|\hat{f}_\infty\left(\Delta\right) \right|^2=q,\nonumber
\end{align}
where the parity of $f\left(t\right)$ was used. The degree of the intrinsic decoherence is thus equal to the relative spin orientation efficiency. Since both are proportional to $\theta^2$, one obviously cannot avoid the loss of coherence by a pulse power adjustment. While unitary spin control may lead to a growth of coherence, the discussed initialization procedure always comes with an inherent decoherence.

%###################################################################
\section{Phonon-induced dephasing}\label{Phonon}
Assuming the pulse area to be small, i.e., the angle of rotation of the state in the Hilbert space induced by it is small, we treat the interaction $V$ given by \eqref{h_las_transform} perturbatively taking the pulse area $\theta$ as a small parameter. As we are interested in the phonon-induced decoherence of the resident hole spins we investigate how the corresponding off-diagonal density matrix element changes due to the phonon-induced dephasing. The calculations are carried out in the second order approximation with respect to $V$\cite{Grodecka}.

Assuming the system to be initially in the product state, $\varrho\left(-\infty\right)=\rho_\mathrm{0}\otimes\rho_\mathrm{T}$ (with $\rho_\mathrm{T}$ being the thermal equilibrium distribution of phonon modes), we start with the evolution equation for the full system density matrix $\tilde{\varrho}\left(t\right)$ in the interaction picture with respect to $H_\mathrm{las}$,
\begin{equation}\label{tilderho}
\begin{aligned}[\columnwidth]
\tilde{\varrho}\left(t\right)\simeq & \tilde{\varrho}_\mathrm{0}+\frac{1}{i\hbar}\int_{-\infty}^{t}\mathrm{d}\tau \left[V\left(\tau\right),\varrho\left(-\infty\right)\right] \\
&-\frac{1}{\hbar^2}\int_{-\infty}^{t}\mathrm{d}\tau \int_{-\infty}^{\tau}\mathrm{d}\tau{'} \left[V\left(\tau{'}\right), \left[V\left(\tau{''}\right),\varrho\left(-\infty\right)\right]\right]
\end{aligned}
\end{equation}
where $V\left(t\right)$ is the interaction Hamiltonian in the interaction picture with respect to $H_\mathrm{las}$.

Taking the trace over the phonon reservoir degrees of freedom we get the reduced density matrix of the carrier subsystem in the Schr{\"{o}}dinger picture
\be
\rho\left(t\right)= U\left(t\right)\mathrm{Tr_R}\tilde{\varrho}\left(t\right)U^{\dagger}\left(t\right)\mathrm{,}\nonumber
\ee
where $U\left(t\right)$ is the evolution generated by $H_\mathrm{las}$.

Upon substituting for $\tilde{\varrho}\left(t\right)$ from Eq.~\eqref{tilderho}, the first term of $\rho\left(t\right)$, $\rho^{(0)}\left(t\right)=U\left(t\right)\varrho_0 U^\dagger\left(t\right)$, obviously gives the unperturbed evolution, while the second one vanishes as it contains the thermal average of an odd number of phonons. Finally, the last term is the perturbative second order correction to the reduced density matrix. After substituting $V$ from Eq.~\eqref{h_las_transform} into Eq.~\eqref{tilderho} and defining the frequency dependent operators
\be\label{y}
Y\left(\omega\right)= \frac{1}{\hbar}\int_{-\infty}^{\infty}\mathrm{d}tS\left(t\right)e^{i\omega t}
\ee
the expression for $\rho$ after the laser pulse takes the form
\be
\rho\left(t\right)=U\left(t\right)\rho_{0}U^{\dagger}\left(t\right)+\Delta\rho\mathrm{,}\nonumber
\ee
with the phonon-induced correction
\begin{equation}\label{poprawka}
\begin{aligned}[\columnwidth]
\Delta\rho=\int_{-\infty}^{\infty}{\frac{\mathrm{d\omega}}{\omega^2}R\left(\omega\right)}&U\left(\infty\right) \left( Y\left(\omega\right)\rho_0 Y^\dagger \left(\omega\right)\vphantom{\frac{1}{2}}\right.\\
&\left.-\frac{1}{2}\left\{ Y^\dagger \left(\omega\right)Y\left(\omega\right),\rho_0 \right\} \right)U^{\dagger}\left(\infty\right)\mathrm{,}
\end{aligned}
\end{equation}
where
\begin{align}\label{spektralna}
R\left(\omega\right)&=\frac{1}{\hbar^2}\sumk{\left|\fk\right|}^2\left|n_B\left(\omega\right)+1\right|\delta\left(\left|\omega\right|-\wk \right)\nonumber
\end{align}
is the phonon spectral density.

Using Eq.~\eqref{y} with the explicit form for $S$ from Eq.~\eqref{s} one finds
\be
Y=\frac{i}{2}\hat{f}_\infty\left(\omega-\Delta\right) \ket{T\!\uparrow}\!\!\bra{\uparrow}+\mathrm{h.c.}\mathrm{,}
\ee
where $\hat{f}_t\left(\omega-\Delta\right)$ is the pulse spectrum shifted by the detuning. 
Substituting this to Eq.~\eqref{poprawka} and keeping only terms up to the second order in $\theta$ we arrive at the solution for the spin coherence after the pulse
\begin{align}\label{exp}
\bra{\downarrow}\rho\left(\infty\right)\ket{\uparrow}&=\bra{\downarrow}U\left(\infty\right)\rho_0 U^{\dagger}\left(\infty\right)+\Delta\rho\ket{\uparrow}\nonumber\\
&\simeq\left(1-w\right)\bra{\downarrow}\rho^{\left(0\right)}\left(\infty\right)\ket{\uparrow} \mathrm{,}\nonumber
\end{align}
where %$\rho_\infty^{\left(0\right)}$ is the unperturbed final density matrix and 
\be\label{w}
w\left(\Delta\right)={\frac{1}{8}} \int_{-\infty}^\infty{\mathrm{d}\omega{\frac{R\left(\omega\right)}{\omega^2}} {\left| \hat{f}_{\infty}\left( \omega-\Delta\right) \right|}^2}
\ee
can be interpreted as the degree of the phonon-induced dephasing. Thus the degree of dephasing is proportional to the overlap between the phonon spectral density and the Gaussian power spectrum of the laser pulse shifted in the frequency domain by the detuning. According to Eq.~\eqref{w}, the phonon-induced dephasing scales with the square of the pulse amplitude, $\theta^2$. This seems intuitively reasonable if one considers the growing efficiency of phonon generation by the driving field\cite{Vagov}. Clearly, the degree of dephasing is correlated with the amount of perturbation in the environment (here the generation of non-thermal phonons) via the "which path" information transfer\cite{Roszak}.

In what follows, we focus on $w$ as a measure of spin-mediated dephasing in the system and investigate its dependence on the laser pulse detuning, $w\left(\Delta\right)$, as compared to the efficiency of the spin initialization and to the intrinsic contribution to dephasing for various system parameters covering both self-assembled and interface fluctuation QD systems.

In the numerical calculation of $w$, the electron and hole wave functions were modeled by Gaussians
\be
\psi_\mathrm{e/h}\left(\bm{r}\right)= \pi^{-\frac{3}{4}}\left(l_\mathrm{z}l_\mathrm{e/h}\right)^{-1} e^{-\frac{1}{2}\big(\frac{r_\perp}{l_\mathrm{e/h}}\big)^2- \frac{1}{2}\left(\frac{z_{\mathrm{e/h}}}{l_\mathrm{z}}\right)^2}\nonumber
\ee
with $l_\mathrm{z}$ and $l_\mathrm{e/h}$ being the wave function localization widths in the growth direction and in the sample plane for electron and hole respectively. We introduce the parameter $d=\left|z_\mathrm{e}-z_\mathrm{h}\right|$, which is used as a measure of wave functions separation in the growth axis.   

%###############################################################
\section{Results}\label{results}
As we have seen, the intrinsic loss of coherence in the optical orientation process is equal to the achieved orientation effect. The phonon-induced contribution is therefore of more interest. Here, we present the results on the phonon-induced dephasing as a function of the system parameters and orientation conditions and then compare them to the intrinsic contribution and orientation efficiency. 

\begin{figure}[tb]
	\begin{center}
		\includegraphics[width=\columnwidth]{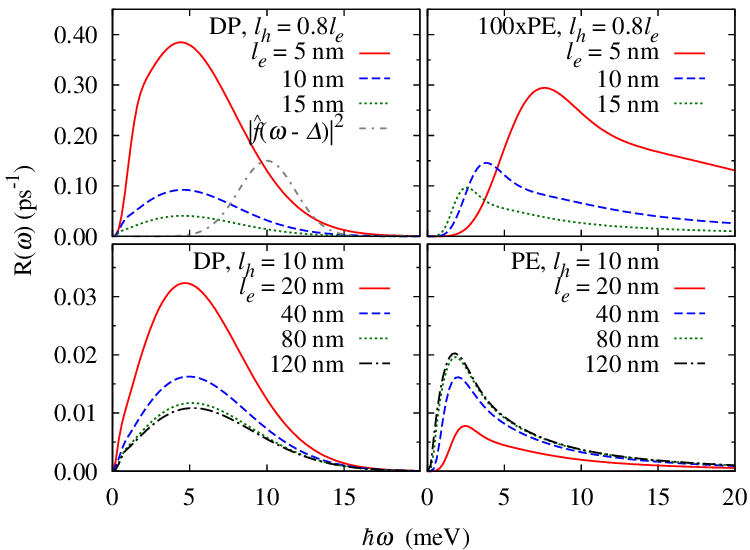}\upemhf
	\end{center}
	\caption{\label{fig:fig1}(Color online) The phonon spectral density, $R\left(\omega\right)$, with its DP (left panels) and PE (right panels) contributions for exemplary systems: three sizes of a self-assembled QD (upper panels) and fluctuation QD with various electron spread volumes (bottom panels). In the upper left panel, the pulse power spectrum $|\hat{f}_\infty\left(\omega-\Delta\right)|^2$ is plotted with grey dash-dotted line for an arbitrary pulse area $\theta$ and  detuning $\hbar\Delta=10~\mathrm{meV}$, at $T=1~\mathrm{K}$.}
\end{figure}
The phonon spectral density is shown in Fig.~\ref{fig:fig1} for two typical systems imitating a self-assembled QD (similar, relatively strong confinement for electrons and holes, $l_\mathrm{e}{\approx}l_\mathrm{h}$) and a fluctuation one (electron weakly confined, $l_\mathrm{e}{\gg}l_\mathrm{h}$) in the upper and lower panels, respectively. The DP and PE contributions are shown separately. Additionally, we plot the pulse power spectrum (grey dash-dotted line). For self-assembled QDs, the PE interaction occurs to be negligible, while it can give a significant contribution in fluctuation QDs with the maximum coupling for phonons at $\omega\approx{2}~\mathrm{meV}$. The DP coupling has its maximum for $\omega\approx{5}~\mathrm{meV}$. The pulse power spectrum $|\hat{f}_\infty\left(\omega\right)|^2$ is much narrower than the spectral density so one can expect that $w\left(\Delta\right)$ given by \eqref{w} will resemble the shape of $R\left(\omega\right)$. Since the phonon-induced dephasing is proportional to $\theta^2$, in the following discussion we will plot the value of $w/\theta^2$, which is independent of the pulse area.

\begin{figure}[tb]
\begin{center}
\includegraphics[width=\columnwidth]{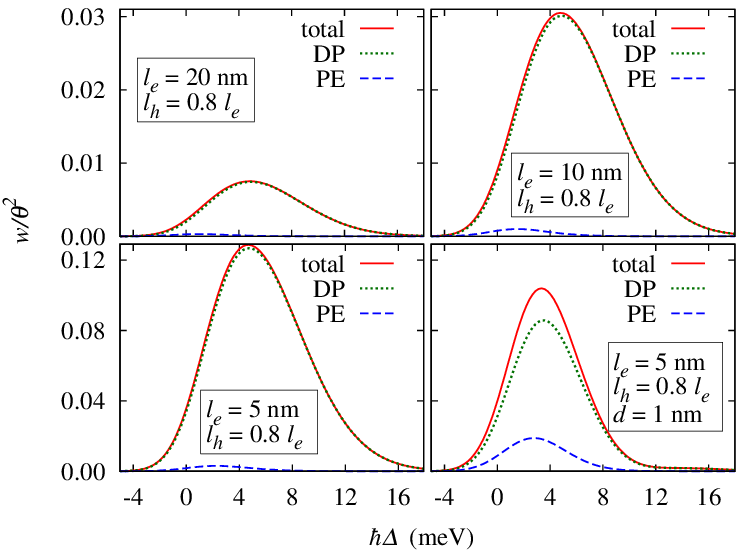}\upemhf
\end{center}
\caption{\label{fig:fig2}(Color online) The deformation potential (dotted green line) and piezoelectric (dashed blue line) coupling contributions to the phonon-induced dephasing as a function of the laser detuning for a different in-plane wave function localization widths with $l_\mathrm{h}=0.8~l_\mathrm{e}$  imitating carriers well confined in a self-assembled QD. In the bottom right panel the wave function separation in the growth axis set to $d=1~\mathrm{nm}$. The solid red line is the sum of the two contributions. Calculations for $T=1~\mathrm{K}$.}
\end{figure}
\begin{figure}[tb]
\begin{center}
\includegraphics[width=\columnwidth]{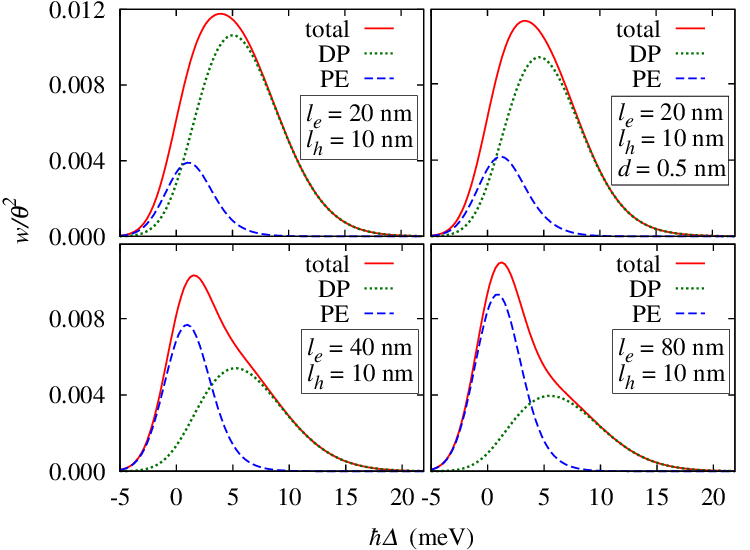}\upemhf
\end{center}
\caption{\label{fig:fig3}(Color online) The deformation potential (dotted green line) and piezoelectric (dashed blue line) coupling contributions to the phonon-induced dephasing as a function of the laser detuning for the hole wave function size $l_\mathrm{h}=10~\mathrm{nm}$ but with varying electron confinement for an interface fluctuation QD. The solid red line is the sum of the two contributions. In the top right panel the wave function separation in the growth axis is set to $d=0.5~\mathrm{nm}$. All calculations for $T=1~\mathrm{K}$.}
\end{figure}
The resulting degree of phonon-induced dephasing for two sets of system parameters that correspond to self-assembled and interface fluctuation QDs is shown in Fig.~\ref{fig:fig2} and Fig.~\ref{fig:fig3}, respectively. One can observe strong dependence of the degree of dephasing on the laser detuning. In all the cases, at low temperture ($T=1~\mathrm{K}$), the decoherence is suppressed by the negative detuning larger than about $5~\mev$ and approximately two times greater positive one.  
In Fig.~\ref{fig:fig2}, we show separately plotted contributions from the piezoelectric and deformation potential couplings to the global effect for a few hole and electron wave function localization widths with a set $l_\mathrm{h}/l_\mathrm{e}$ ratio equal $0.8$. This corresponds to a self-assembled QD system, where both carriers are well confined in comparable volumes. One can notice the strong dependence of the effect on the QD size: the more the carriers are confined the stronger the dephasing, reaching several percent for the confinement typical for, e.g., a self-assembled InAs/GaAs and detunings of a few meV, resonant with the effectively most strongly coupled phonons (see Fig.~\ref{fig:fig1}). The lower right panel shows that a small carrier separation along the growth axis, $d=1~\mathrm{nm}$, lowers the total dephasing value due to reduced DP contribution while an increase of the PE part is observed.

Fig.~\ref{fig:fig3} shows the degree of phonon-mediated dephasing for the interface fluctuation QDs. We start with a result for a sample in which the hole and electron wave functions are spread over a comparable volume (both types of carriers confined inside a large QD) and then gradually stretch the electron wave function in the in-plane direction to simulate the natural QD system with heavy holes strongly confined on the potential fluctuation and electrons loosely bound to them. With this transition, the competition between the contributions from DP and PE couplings takes place with the PE interaction dominating for natural QDs ($l_e\gg l_h$) while the total value of the degree of dephasing declines. This can easily be understood because the PE interaction grows with the carrier spatial displacement that suppresses charge compensation. Overall, for a fluctuation QD, the PE coupling contributes much more to the phonon ultrafast dephasing than in the case of a self-assembled dot. The drop of the total effect for fluctuation QDs is connected with the raising spread of carrier wave functions. Additionally, the effect of a small, $d=0.5~\mathrm{nm}$, displacement between electron and hole wave functions along the growth axis is studied with its result shown in the top right panel. One can notice a small rise of the PE contribution due to a further charge separation.  

\begin{figure}[tb]
\begin{center}
\includegraphics[width=\columnwidth]{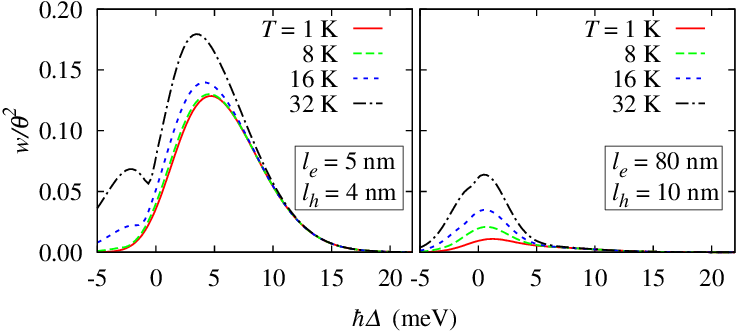}\upemhf
\end{center}
\caption{\label{fig:fig4}(Color online) The dependence of the total phonon-induced dephasing on the temperature for two sets of parameters reflecting a typical self-assembled QD (left panel) and a fluctuation QD (right panel).}
\end{figure}
Fig.~\ref{fig:fig4} presents how the phonon-induced dephasing scales with temperature. For both sets of parameters (imitating carriers confined in a self-assembled QD and the natural QD system), one can observe some increase of the degree of dephasing as the temperature grows. The effect of temperature is relatively larger in the case of fluctuation-like QDs. For systems of both types the phonon-induced dephasing is enhanced by the temperature (switching on phonon absorption) more on the negative detuning side, while its slope far on the positive side remains unchanged. In the case of a self-assembled QD at $T=1~\mathrm{K}$ the maximum of the degree of dephasing is at roughly $5~\mathrm{meV}$ detuning. For the resonant excitation one gets approximately $4\%$ degree of dephasing and a small negative detuning of about $2~\mathrm{meV}$ is needed to reduce it below the level of $1\%$. At higher temperatures the maximum shifts slightly and, which is more interesting, a secondary peak appears on the negative detuning side. Having in mind that the orientation efficiency drops with detuning, one finds the local minimum at less than $1~\mathrm{meV}$ negative detuning to be a candidate for the optimal excitation condition at higher temperatures. For a fluctuation QD, the dependence of the degree of dephasing becomes more symmetrical with rising temperature. The maximal value, roughly equal to the value for the resonant excitation, shifts even more towards zero detuning and varies from about $1\%$ at $T=1~\mathrm{K}$ to above $5\%$ at $T=32~\mathrm{K}$.

The results presented above show that the phono-induced dephasing decreases for detuned pulses which might suggest that increasing the detuning is a way to retain more coherence while orientationg the spins. It is clear, however, that the efficiency of spin orientation also decreases for detuned pulses. A reasonable figure of merit is the ratio of the phonon-induced loss of coherence, $w$, and the orientation efficiency $q$. Since both $w$ and $q$ are proportional to $\theta^2$ this quantity does not depend on the pulse area. Moreover, since the relative intrinsic dephasing is equal to $q$ the ration $w/q$ simultaneously yields information on the relative importance of the two dephasing channels.

\begin{figure}[tb]
	\begin{center}
		\includegraphics[width=\columnwidth]{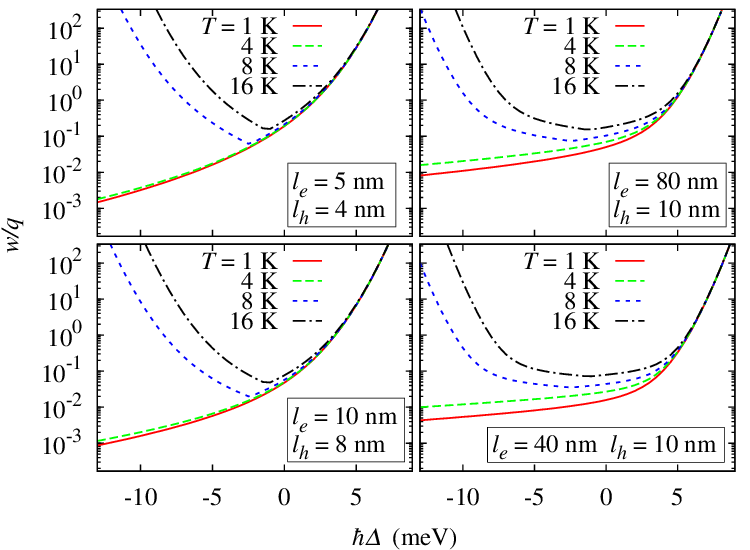}\upemhf
	\end{center}
	\caption{\label{fig:fig5}(Color online) The ratio of the degree of phonon-induced dephasing to the initialization efficiency, $w/q$, as a function of the laser detuning for various temperatures. The left panels contain the plots for two sizes of a self-assembled QD. In the right panels a fluctuation QD.}
\end{figure}
Fig.~\ref{fig:fig5} presents the dependence of $w/q$ on the laser detuning for different sizes of both self-assembled and fluctuation QDs at various temperatures. For systems of both types at low temperature the negative detuning is always favorable in view of the decreasing $w/q$ ratio. The maximal value of detuning to be used can be estimated from the dependence of $q$ on the detuning for a desired efficiency level. With the rise of temperature there is a qualitative change: a minimum of $w/q$ appears at a low negative detuning, yielding the condition for an optimal excitation. It is placed at about $-2.5~\mathrm{meV}$ at $T=8~\mathrm{K}$ and is shifted towards zero detuning with rising temperature. The $w/q$ values within $\left(0.1-0.01\right)$ are achievable at this temperature. For the fluctuation QDs the slope in vicinity of the minimum is small, therefore in such systems the condition for the optimal excitation is less strict.   

%#########################################################################
\section{Conclusions}\label{conclusions}
We have investigated the recently proposed spin initialization scheme\cite{Saper, Kugler, Studer} from the point of view of its efficiency and the amount of dephasing unavoidably built into the process. We have identified two channels of spin dephasing: one intrinsic, related to the initialization scheme itself, and other one resulting from coupling to phonons. Both contributions are proportional to the optical pulse power. Since the efficiency of spin orientation shows the same dependence, there is a fixed proportionality relation between the achieved orientation effect and the incurred dephasing, thus the loss of coherence cannot be effectively reduced by using weak excitation. The decoherence caused by the excitation process itself is equal to the obtained optical spin orientation efficiency and drops down with the increase of the laser detuning, equally for positive and negative. The phonon-induced dephasing dominates over the latter in the positive detuning range (greater that a few $\mathrm{meVs}$) at low temperatures (a few $\mathrm{K})$. At higher temperatures there is a closed range of detuning values shifted to the negative side for which the phonon-induced effect is weaker. 

The investigation of the ratio of the phonon-induced dephasing to the spin orientation efficiency yields the conditions for an optimal optical excitation. At low temperatures (a few $\mathrm{K}$) the negative detuning is always advantageous. With the rise of temperature a strict condition for the minimal value of $w/g$ arises making the small negative detuning (a few $\mathrm{meVs}$) most favourable.

While the pulse-driven dephasing is independent of the sample details, the strength of the phonon-induced dephasing depends critically on the confinement size of the carriers as well as on the displacement between the electron and the hole. Our results are applicable both to self-assembled QD systems and natural (fluctuation QD) ones where the phonon-induced effect is less important. The degree of dephasing associated with reasonable polarization efficiency via the discussed initialization method is at least on the order of a few percent, which is large from the point of view of quantum computing applications. The occurrence of a comparable degree of such an ultrafast hole spin dephasing in similar system have been shown experimentally\cite{Kugler}.

While our considerations referred to a p-doped system, motivated by recent experiments\cite{Saper, Kugler, Studer}, all the derivations and statements about the dynamical dephasing made by us are applicable also for systems with resident electrons\cite{Shabaev, Yugova, GlazovReview}. The excitation process, regardless of the type of resident charge present in the system, consists of creating an additional electron-hole pair hence the phonon-induced dephasing effect is symmetric with respect to the type of the resident carrier.

%############################################################################
\section*{Acknowledgements}
We would like to thank Tobias Korn and Michael Kugler for discussions.
This work was partly supported by the TEAM programme of the Foundation for Polish Science, co-financed by the European Regional Development Fund.


%\bibliography{qw}
\end{document}